\documentclass[aps,preprint]{revtex4}
\usepackage{amsmath}
\usepackage{epsfig}
\usepackage{float}
\usepackage{dcolumn} 

\DeclareMathAlphabet{\mathpzc}{OT1}{pzc}{m}{it}
\newcommand{\be}{\begin{equation}}
\newcommand{\ee}{\end{equation}}
\newcommand{\bea}{\begin{eqnarray}}
\newcommand{\eea}{\end{eqnarray}}
\newcommand{\beas}{\begin{eqnarray*}}
\newcommand{\eeas}{\end{eqnarray*}}

\begin{document}
\title{\bf PT-symmetric potentials with imaginary asymptotic saturation}  
\author{Zafar Ahmed$^{1,*}$ and Sachin Kumar$^2$}
\affiliation{$~^1$Nuclear Physics Division,  Bhabha Atomic Research Centre, Mumbai 400 085, India\\
	$~^*$Homi Bhabha National Institute, Mumbai 400 094 , India	\\
	$~^2$Theoretical Physics Section, Bhabha Atomic Research Centre, Mumbai 400 085, India}
	\email{1:zahmed@barc.gov.in, 2: sachinv@barc.gov.in}   
\date{\today}

\begin{abstract}
We point out  that PT-symmetric potentials $V_{PT}(x)$ having imaginary asymptotic 
saturation: $V_{PT}(x=\pm \infty) =\pm i V_1, V_1 \in \Re$ are devoid of scattering states
and  spectral singularity. We show the existence of real (positive and negative) discrete spectrum both with and without complex conjugate pair(s) of eigenvalues (CCPEs). If the states are arranged in the ascending order or real part of discrete eigenvalues,  the initial states  have few nodes but latter ones oscillate fast. Both real and imaginary parts of $\psi(x)$ vanish asymptotically, $|\psi(x)|$ for the CCPEs  are asymmetric and for real energies these are symmetric about origin. For CCPEs $E_{\pm}$ the eigenstates $\psi_{\pm}$ follow an interesting property that $|\psi_+(x)|= N |\psi_-(-x)|, N \in \Re^+$. We remark that, the fast oscillating real discrete energy states discussed are likely to be confused with: reflectionless states, one dimensional version of von Neumann states of Hermitian and spectral singularity state of complex PT-symmetric potentials.
\end{abstract}
\maketitle
Complex PT-symmetric potentials $V_{PT}(x)$ which are  divergent function of $x$ are known to have real discrete spectrum below or above exceptional points $g_n$ [1,4]. Complex PT-symmetric potentials which are asymptotically convergent are not exceptional in this regard [5,7]. The first $V_{PT}(x)$ with imaginary asymptotic saturation was proposed and studied in the form of complex Rosen-Morse potential [6,8], real discrete spectrum has been revealed which is null in the absence of 
the real attractive part of the potential.  More recently, the Dirac delta  potential with a simple imaginary step  added to it, is found to have one real discrete eigenvalue. A new model:  a square well potential with imaginary saturation has been solved and shown to degenerate to the result of Dirac delta model in limit $a \rightarrow 0$ [9]. However, two contradictory ideas have been professed: in Rosen-Morse case the real discrete spectrum comes from the poles [8] of transmission amplitude $t(E)$, but in the square well and the  Dirac well case the discrete eigenvalue is obtained as a zero [9] of the reflection amplitude $r(E)$. 

In this work, we bring attention to the fact that when a potential has imaginary asymptotic saturation $V_{PT}(\pm \infty)=iV_1, V_1 \in \Re$, the propagating plane waves $e^{\pm ikx}$,
$k=\sqrt{2m E}/\hbar$ can no more be solutions of the Schr{\"o}dinger equation even in the asymptotic region. One will instead have $\psi(x) \sim e^{(\pm \alpha+i\beta)x}$ on the left and right sides of the potential. Their current densities do not equilibrate as they become local (function of $x$). Consequently it is not really possible to define reflection $r(E)$ and transmission $t(E)$ amplitudes. Seeing the discrete spectrum sometime as poles of $t(E)$ and other times as zeros of $r(E)$ has been ad-hoc [8,9]. Moreover, reflection being non-reciprocal [10] for left and right injection,  one may also get two sets of (non-unique) discrete spectrum for such PT-symmetric potentials. 

In this paper, we re-examine PT-symmetric potentials (see Eqs. (3,8,11,16) below) with imaginary asymptotic saturation using new solvable and two existing  [6,8,9] potential models. We bring out their new crucial features which have been missed out earlier. 

The Schr{\"o}dinger equation for a general potential is written as
\begin{equation}
\frac{d^2\psi(x)}{dx^2}+\frac{2m}{\hbar^2}[E-V(x)]\psi(x)=0.
\end{equation}
In the following we solve (1) for various PT-symmetric potentials with imaginary asymptotic saturation to obtain two kinds of eigenvalues- complex conjugate pairs or real (positive and negative) discrete eigenvalue or both. These potentials cannot have  continuum of real eigenvalues and spectral singularity. In general the boundary condition to be imposed of eigenfunctions is
\begin{equation}
\psi(x \sim -\infty){=} A e^{K_1 x},~ \psi(x \sim \infty){=}B e^{-K_2 x},~K_1{=}\sqrt{-\frac{2m}{\hbar^2}(E+iV_1)},~  K_2{=}\sqrt{-\frac{2m}{\hbar^2}(E-iV_1)}.
\end{equation}
Alternatively, if we define $K_1=k_1+ik_2$ and $K_2=k_1-ik_2$, where $k_1,k_2 >0$, we see decaying solutions both the sides. On the left $\psi(x\sim -\infty)\sim A e^{(k_1+ik_2)x}$ and on the right $\psi(x \sim \infty) \sim B e^{-(k_1-ik_2)x}.$ Interestingly at any energy, all the solutions will be decaying asymptotically. But only  at special discrete energies, they will be continuous and differentiable every where and specially at $x=0$. The definition of $K_1$ and $K_2$ (2) will remain the same for all the potentials to be discussed afterwards.

{\bf 1. Simple step potential:}

First, we see an  absence of  spectrum in the simple imaginary step potential
\begin{equation}
V(x\le 0) =-iV_1, \quad V(x>0)=iV_1,
\end{equation}
as the solutions of (1) for (3) which vanish at $x = \mp \infty$, respectively. Matching the wavefunctions (2) and their derivative at $x=0$, we get a equation $2k_1=0$, which does not have any real or complex root for $V_1 \ne 0$. So no  (discrete or continuous)  energy state is possible in this complex PT-symmetric potential which is two piece.

 {\bf 2. Two piece exponential step:} 
 
This new PT-symmetric potential step can be written as
\begin{equation}
V(x \le 0)=-iV_1[1-e^{2x/a}],\quad V(x > 0)= iV_1[1-e^{-2x/a}].
\end{equation}
Using a transformation $z=e^{|x|/a}$, the Schr{\"o}dinger equation (1) for this potential can be transformed in to a cylindrical Bessel equation to write the solutions of (1) for (4) as,
\begin{equation}
\psi(x\le 0){=}N J_{K_2a}(q) I_{K_1a}(q e^{x/a}), \quad \psi(x>0){=}N I_{K_1a}(q) J_{K_2a}(q e^{-x/a}), \quad q{=}a\frac{\sqrt{2miV_1}}{\hbar}.
\end{equation}
Here $J_{\nu}(z)$ and $I_{\nu}(z)$ are cylindrical Bessel and modified Bessel functions, which for very small values of $z$ vary as $z^\nu$. Consequently, the solutions become asymptotically decaying waves on both sides as in Eq. (2). By matching the derivative of solutions (5) at $x=0$, we get the energy quantization condition as
\begin{equation}
f(E)=J_{K_2 a}(q) I'_{K_1a}(q)+I_{K_1 a}(q) J'_{K_2a}(q)=0.
\end{equation}
This equation is invariant under the exchange of subscripts 1 and 2, so under the reflection of the potential $V(x)$, the roots of (6) which are eigenvalues, remain invariant. 
For various cases of parameters $V_1$ and $a$, we study the contours of  real and complex parts of $f(E)$ in complex energy plane $E={\cal E}+ i\gamma$, to explore the roots of $f(E)$. In Fig. 1, the contour plots  of $\Re(f(E))=0$ and $\Im(f(E))=0$ in complex plane $({\cal E}, \gamma)$ for the case of $V_1=5$ and $a=8$ present no point of intersection at $\gamma=0$ ($x$-axis), hence there is no real discrete eigen value, but there are three CCPEs: $E_{\pm}={\cal E}_n\pm i\gamma_n$. The real and imaginary parts of $\psi_n(x)$ oscillate but vanish asymptotically. We find  an interesting reflection property of the eigenfunctions:
\begin{equation}
|\psi_+(x)|=N ~ |\psi_-(-x)|,\quad N \in \Re^+,
\end{equation} 
is satisfied, see (b) and (c) parts of Fig. 1.
\begin{figure}[h]
	\centering
	\includegraphics[width=6 cm,height=6.cm]{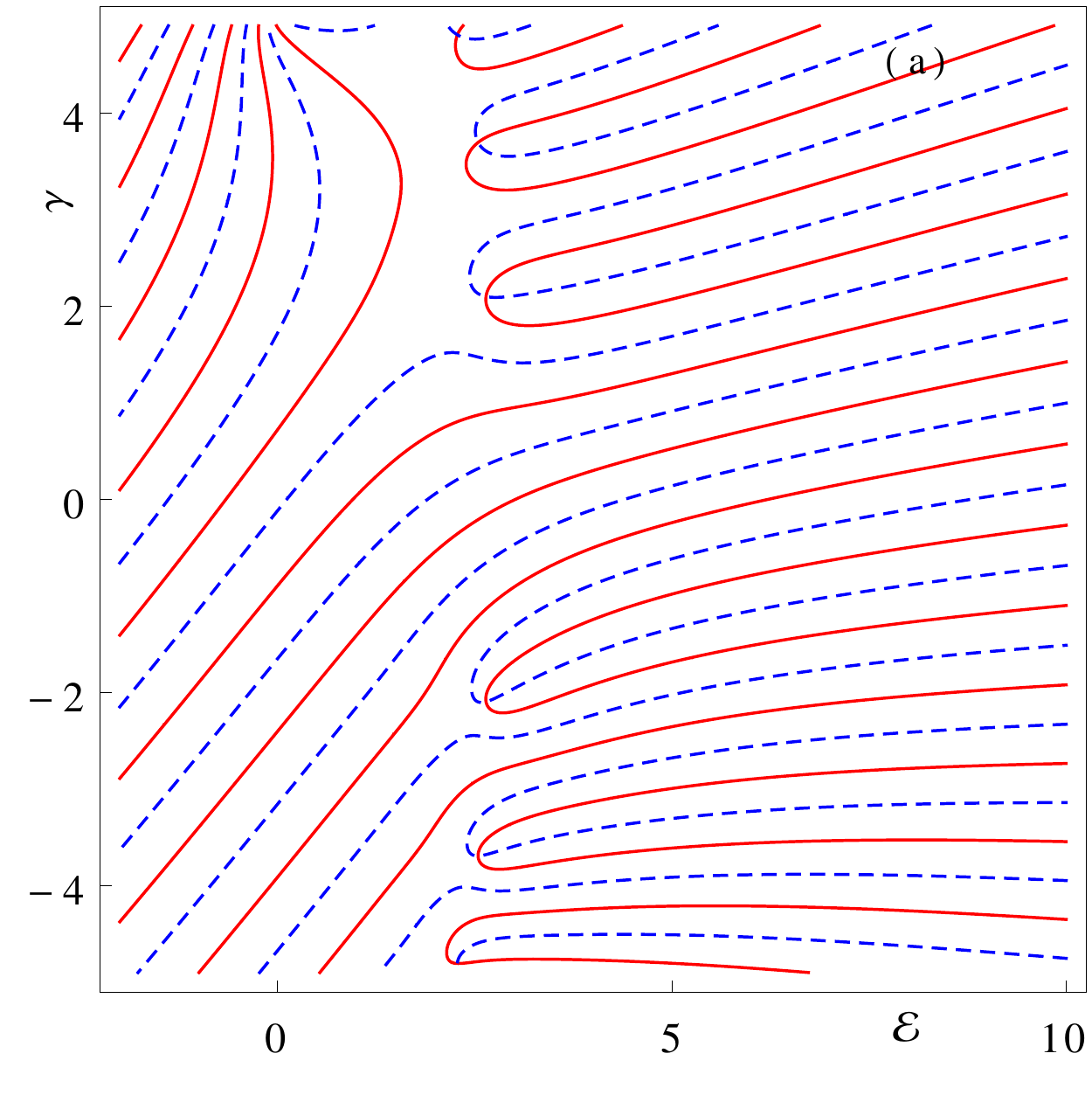}
		\includegraphics[width=5 cm,height=6.cm]{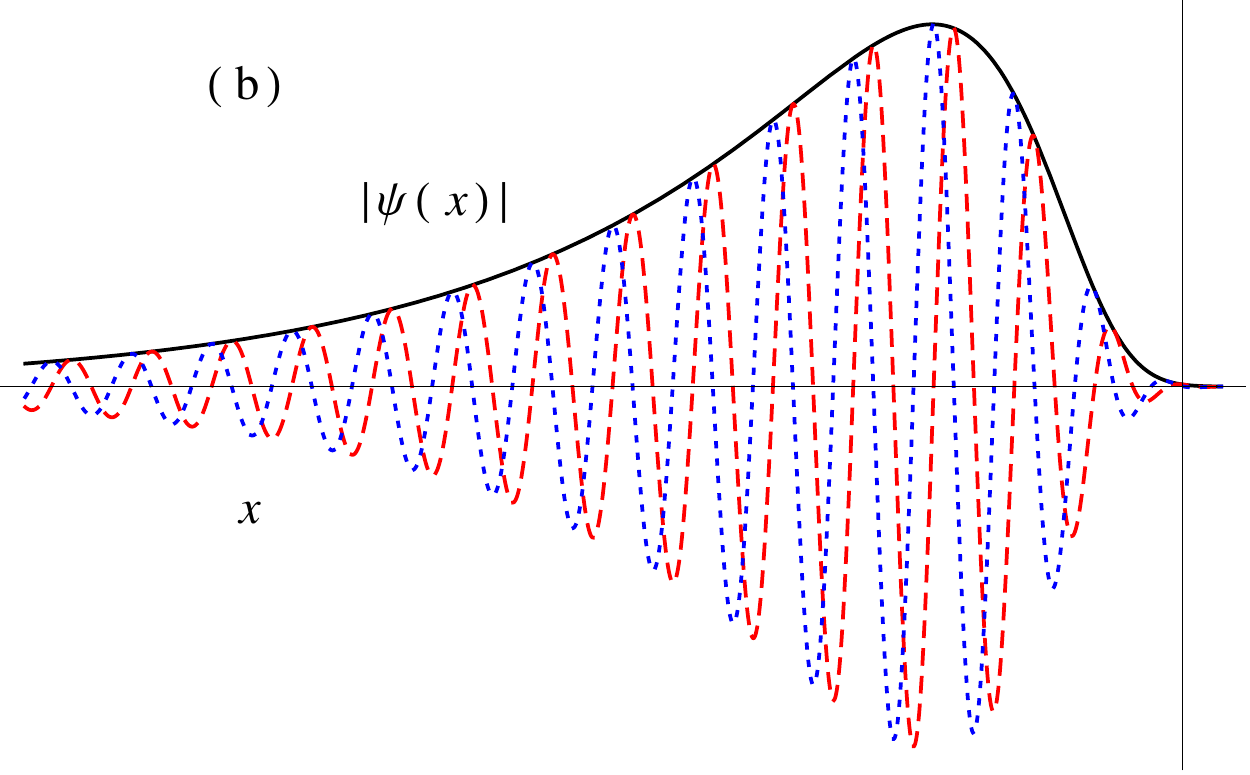}
		\includegraphics[width=5 cm,height=6.cm]{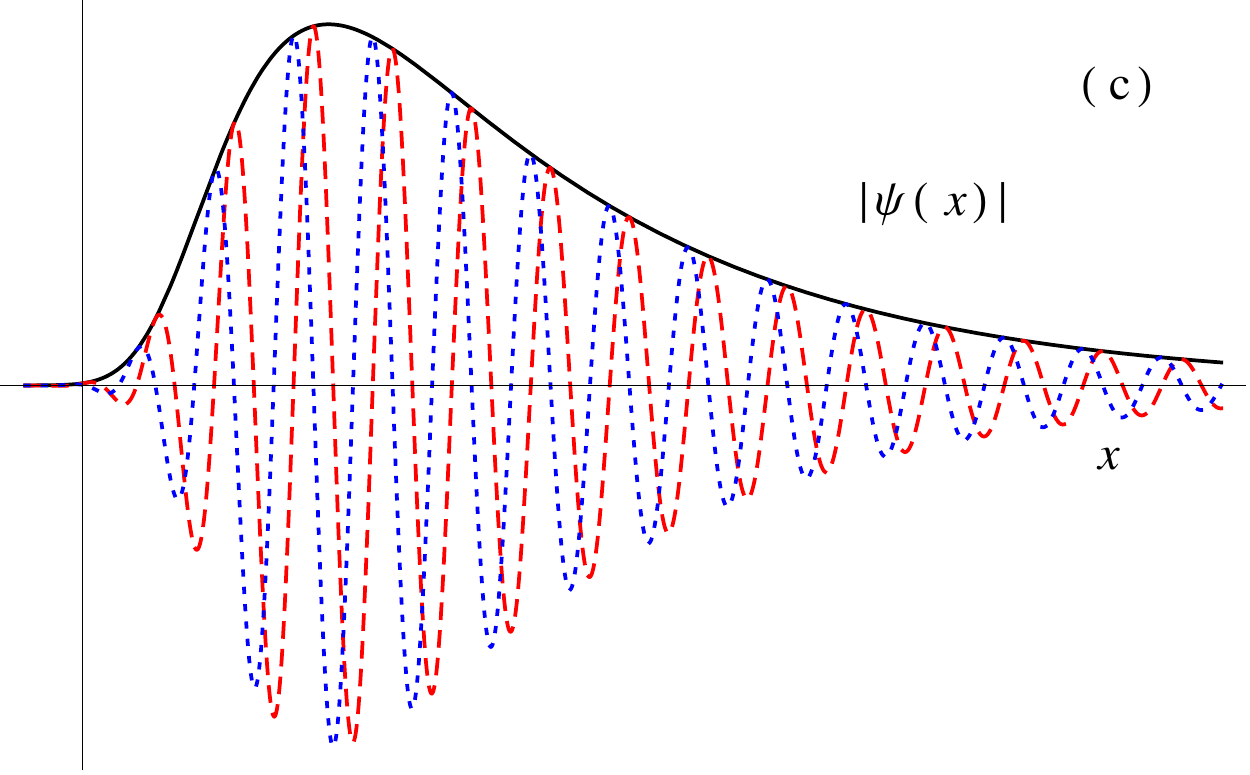}	
 		\caption{Two piece exponential (4) potential for a  case: $V_1=5, a=8$, (a): potential is devoid of real discrete spectrum for this parametric case, it has three CCPEs as there are three points of intersection (corresponding eigen values:  $ E_{1\pm}=2.28\pm 4.80 i, E_{2\pm}=3.54 \pm 3.69 i, $ and $E_{3\pm}=2.64 \pm 2.09$) of the contours $\Re(f(E))=0$ (solid) and $\Im(f(E))=0$ (dashed) in complex energy plane $({\cal E},\gamma)$. Note that there are not more than three points of intersection on either side of y-axis (i.e. $\gamma$-axis), and for clarity of figure we have kept y-axis up to three points of intersection. (b):  $|\psi(x)|$ (solid), $\Re(\psi(x))$ (dashed) and $\Im(\psi(x))$ (dotted) are plotted for $E=E_-$. (c): The same as in (b) for $E=E_+$. $|\psi_{mp}|$  peak on left (right) of $x=0$, this trend will reverse for $V_1=-5$ and $a=8$.}
	\end{figure}

{\bf 3. Linear three piece step:}
Linear three piece PT-symmetric step potential with imaginary asymptotic saturation can be created as
\begin{equation}
V(x \le -a) =-i V_1,\quad  V(-a<x< a) =iV_1x/a, \quad V(x \ge a)= i V_1.
\end{equation}
The solution of Schr{\"o}dinger equation (1) for this potential can be written as
\begin{equation}
\psi(x \le -a)= A e^{K_1x}, \quad \psi(-a< x< a)= C Ai(h(x))+D Bi(h(x)), \quad \psi(x \ge a)= B e^{-K_2x}.
\end{equation}
Here $h(x)=\frac{2m}{g\hbar^2}(E-iV_1\frac{x}{a})$, $g=\left(\frac{2mV_1}{\hbar^2 a}\right)^{2/3}$. $Ai(h)$ and $Bi(h)$ are called Airy functions. Let us introduce $h_{1,2}=(E\pm iV_1)/g$ and $h_0=-i\sqrt{g}$. By matching these solutions and their derivative at $x=\pm a$, we get linear equations in $A,B,C$ and $D$. By eliminating them, we get an eliminant which gives the eigenvalue equation as,
\begin{eqnarray}
f(E)=K_1 K_2[Ai(h_2)Bi(h_1)-Ai(h_1)Bi(h_2)]-iK_1 \sqrt{g}[Ai'(h_2)Bi(h_1)-Ai(h_1)Bi'(h_2)] \nonumber \\
-iK_2 \sqrt{g}[Ai'(h_1)Bi(h_2)-Ai(h_2)Bi'(h_1)]-g[Ai'(h_1)Bi'(h_2)-Ai'(h_2) Bi'(h_1)]=0.
\end{eqnarray}
The invariance of (10) under exchange of subscripts 1 and 2 ensures unique spectrum even if the potential is reflected about $y$-axis. In fig. (2), the contour plots for the case of $V_1=5$ and $a=2$, depict one CCPE and two real positive discrete spectrum which is unbounded from above. $|\psi(x)|$ for CCPE is asymmetric function, while for real discrete energies it is symmetric. At higher eigenvalues, oscillations increase but vanish asymptotically. Eigenstates corresponding to CCPE again follow the invariance property suggested in Eq.(7).
\begin{figure}[h]
	\centering
	\includegraphics[width=6 cm,height=6.cm]{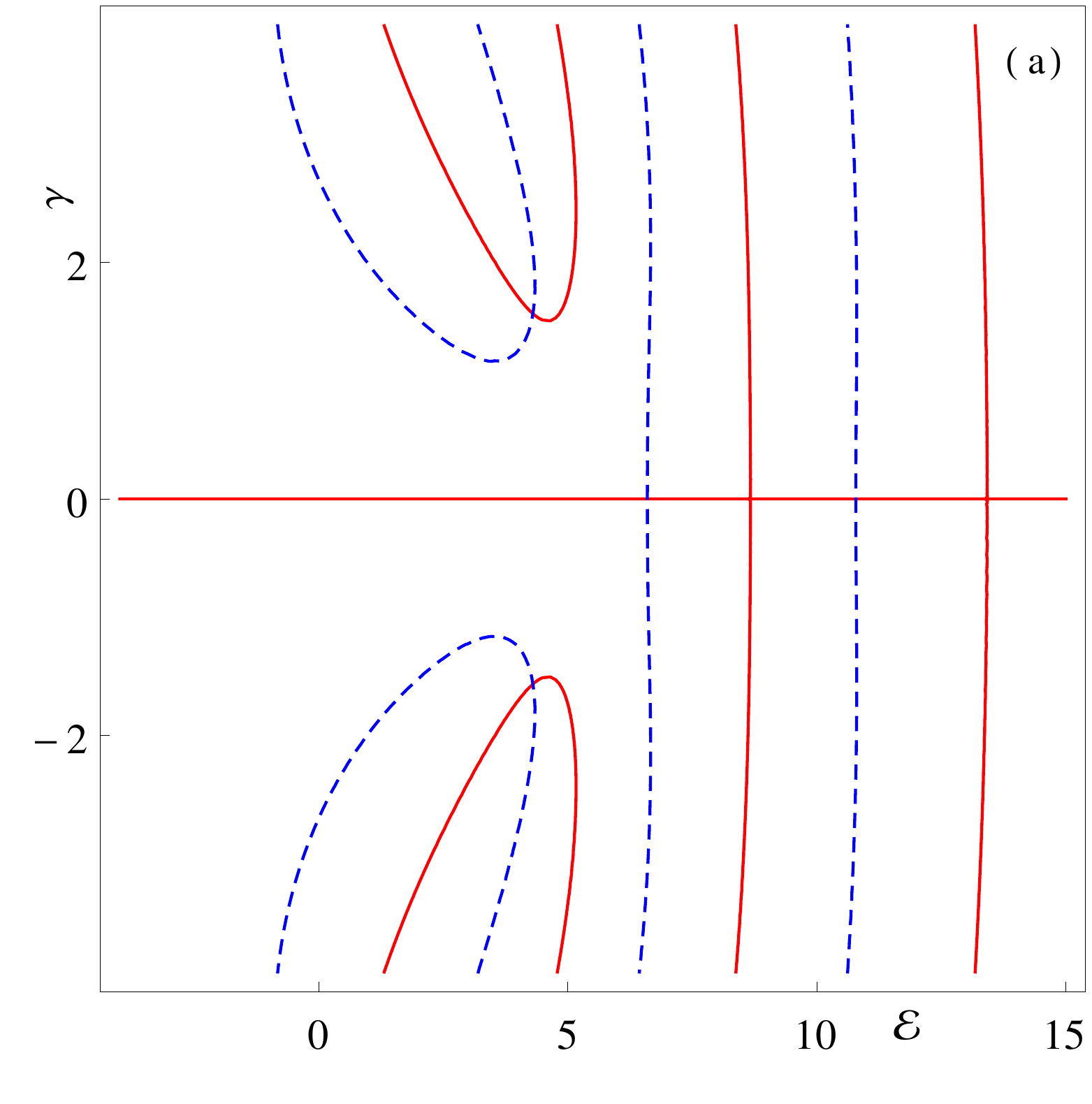}
	\includegraphics[width=5 cm,height=6.cm]{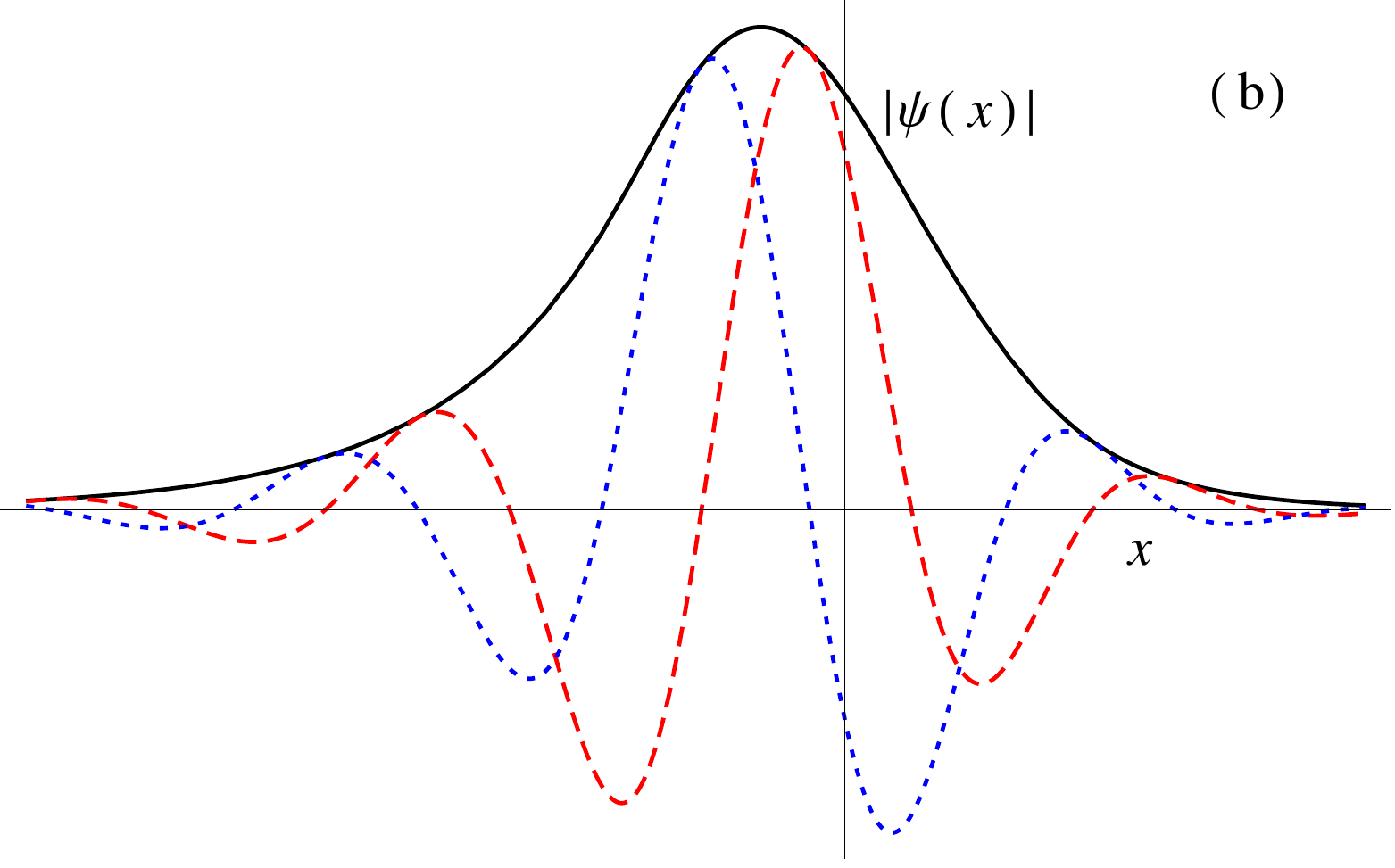}
	\includegraphics[width=5 cm,height=6.cm]{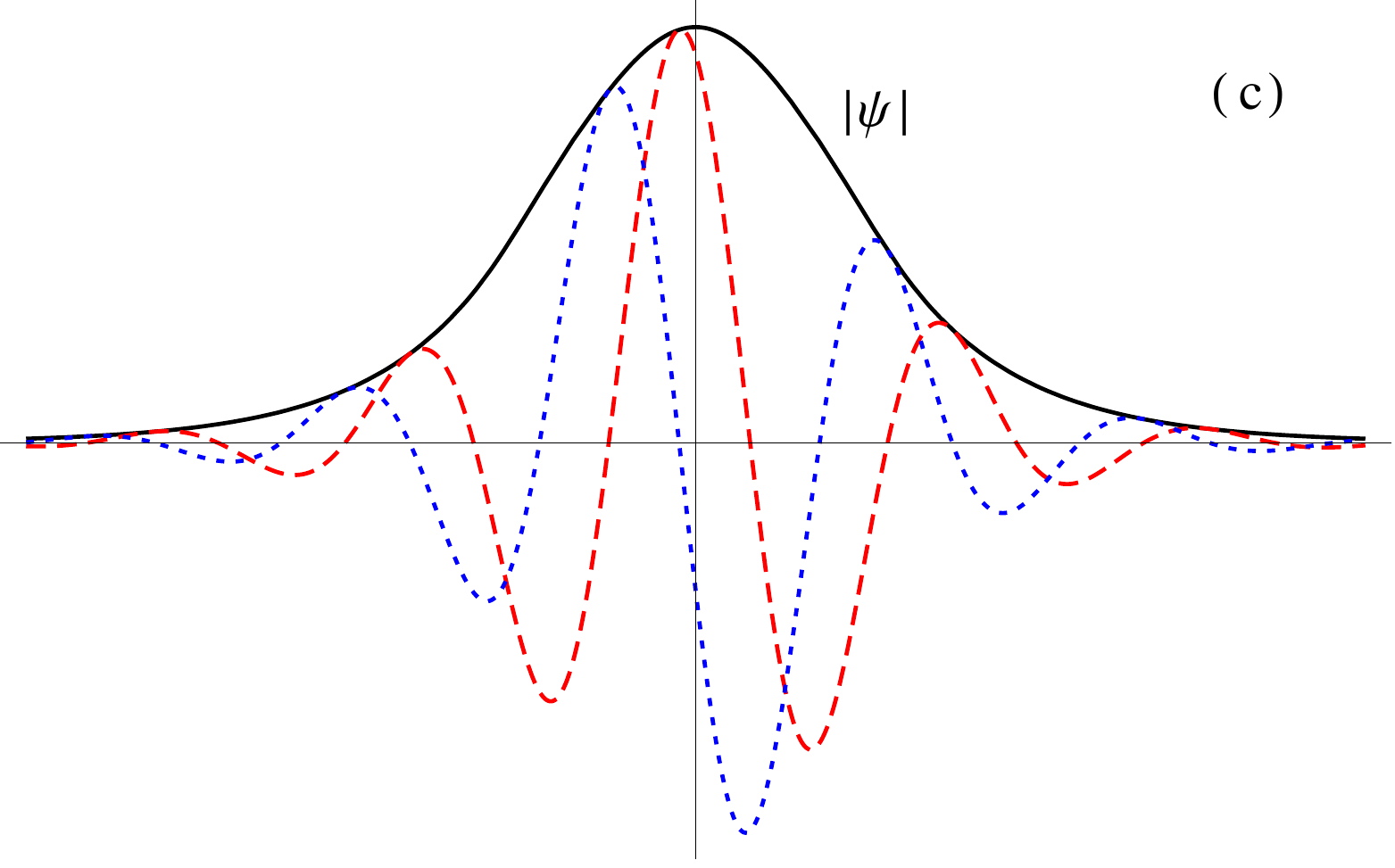}	
	\caption{Linear three piece step (8) potential has both of real discrete spectrum and complex conjugate pairs of eigenvalues, when $V_1=5$ and $a=2$. (a): CCPE ($E_{1\pm}=4.2959\pm 1.5653i,)$ alongwith unbounded real spectrum ($ E_2=6.5952, E_3=10.7814,....$) are plotted here. (b):$|\psi(x)|$ (solid), $\Re(\psi(x))$ (dashed) and $\Im(\psi(x))$ (dotted) are plotted for $E=E_1$, (c): The same as in (b) for $E=E_2$. Notice the for real energy, $|\psi|$ peaks at $x=0$, and for complex energy the peak shifts from $x=0$.}
\end{figure} 

{\bf 4. A general square well potential:}

We can construct a general PT-symmetric square well potential with imaginary asymptotic saturation as
\begin{equation}
V(x{<-}a){=} -iV_1, V({-}a{\le}x\le a){=-V_0}, V(x{\ge}a){=} iV_1,
\end{equation}
It is one of the models discussed in Ref. [9], where only one discrete real energy has been known so far. By defining $p{=}\sqrt{\frac{2m}{\hbar^2}(E+V_0)}$, we can write the full solution of (1) for (11) as
\begin{eqnarray}
\psi(x<-a)=A e^{K_1x}, \psi(-a \le x\le a)= C \sin px +D \cos px,
\psi(x\ge a) = B e^{-K_2 x}~~~
\end{eqnarray}
By matching these solutions and their derivative at $x=\pm a$, we eliminate $A,B,C$ and $D$ to get the energy quantization condition as
\begin{eqnarray}
f(E)=p \cos pa [-(K_1 + K_2) p \cos pa + (-K_1 K_2 + p^2) \sin pa] + \nonumber \\
\sin pa [(-K_1 K_2 + p^2) p \cos pa + (K_2 p^2 + K_1 p^2) \sin pa)]=0.
\end{eqnarray}
which simplifies to
\begin{equation}
f(E)=(K_1+K_2)p \cos 2pa + (K_1 K_2-p^2) \sin 2pa=0.
\end{equation}
This equation is invariant under the exchange of the subscripts 1 and 2.
Further, for the Hermitian square well potential $V_1=0$ $K_1=K_2=K=\sqrt{-2mE}/\hbar$ and $p=P=\sqrt{\frac{2m}{\hbar^2}(E+V_0)}$. In this special case, we recover the well known eigenvalue formulae for even and odd states from (13,14) as
\begin{equation}
\tan 2Pa=\frac{2KP}{P^2-K^2}\quad  \Leftrightarrow \quad  \tan Pa =\frac{K}{P} \quad \mbox{and} \quad \tan Pa=-\frac{P}{K}.
\end{equation}
For the case $V_0=0$, $V_1=5$ and $a=2$, fig. 3 depicts real, positive, discrete energy spectrum of (11) as  the intersections of the contours of real and imaginary parts of $f(E)$ in  Eqs. (13,14). The spectrum is unbounded from above. When $V_1=2, V_0=5$ and $a=2$, see Fig 4, the potential has two CCPEs real discrete positive energies. When $V_1$ is increased the low lying CCPEs start disappearing but the unbounded spectrum of discrete positive energies remains. More interestingly when  we have a square barrier ($V_0=-5$) in between, we get a one CCPE and unbounded discrete spectrum of positive eigenvalues (see Fig. 5).
\begin{figure}[t]
	\centering
	\includegraphics[width=6 cm,height=6.cm]{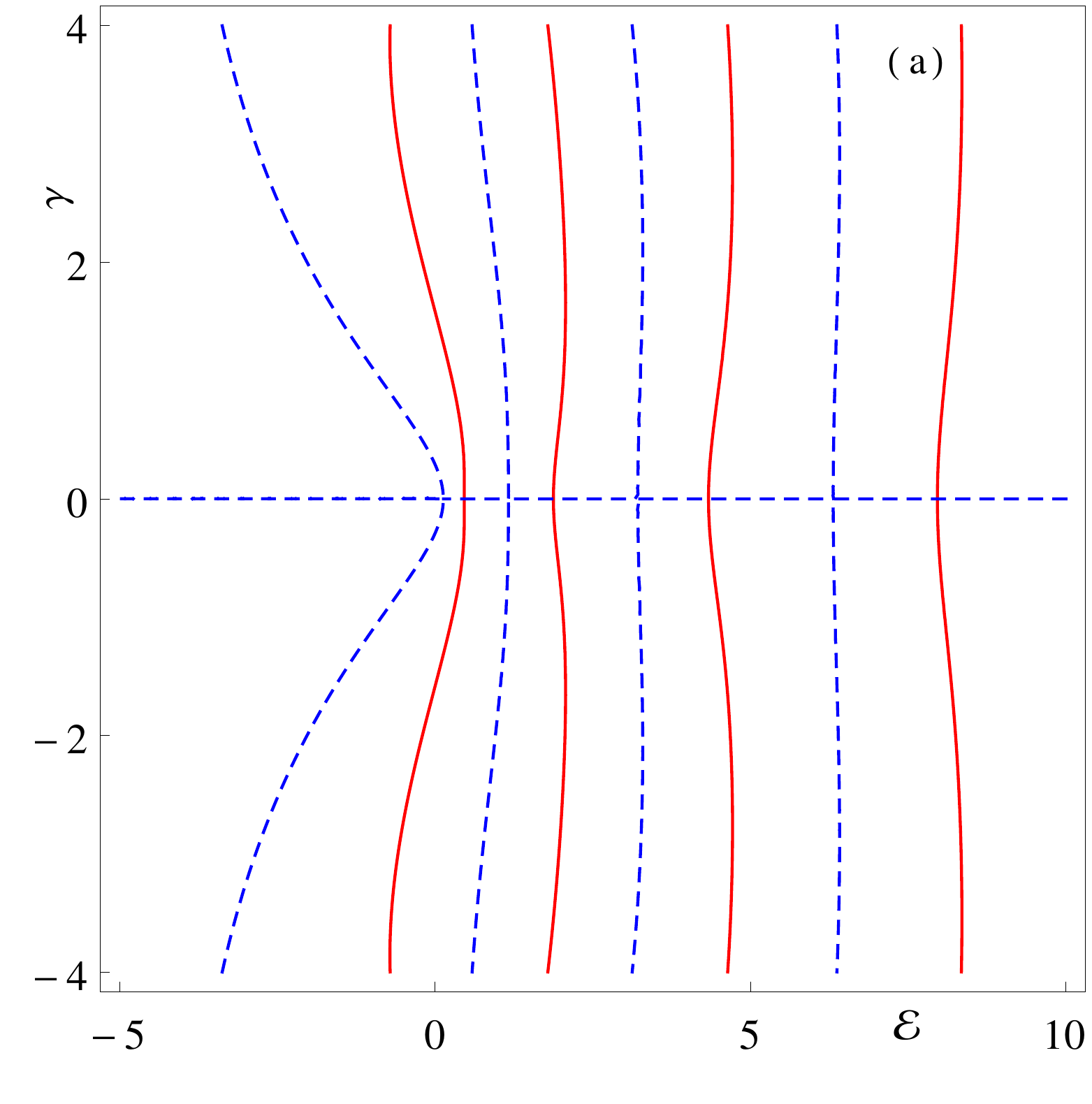}
	\includegraphics[width=5 cm,height=6.cm]{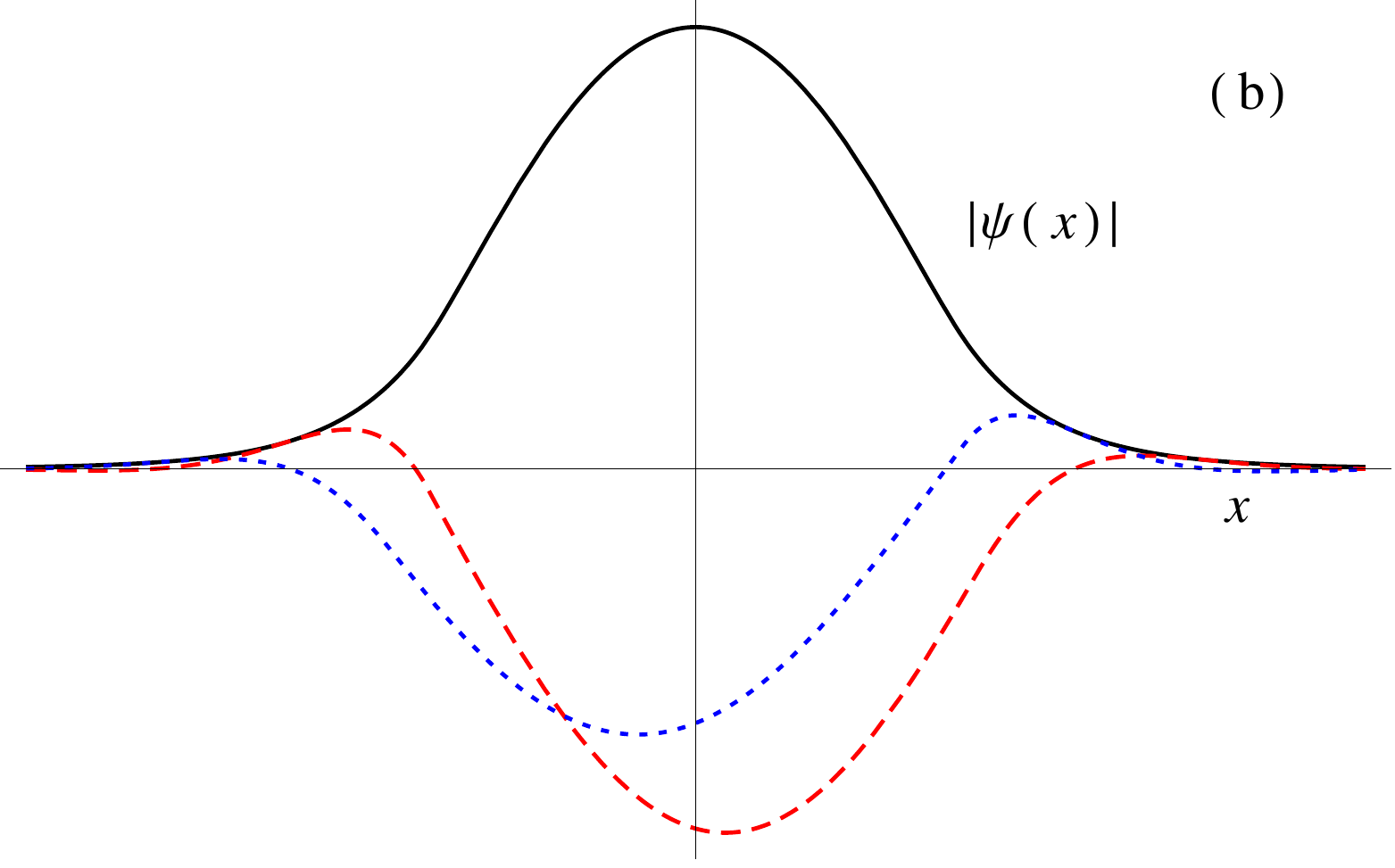}
	\includegraphics[width=5 cm,height=6.cm]{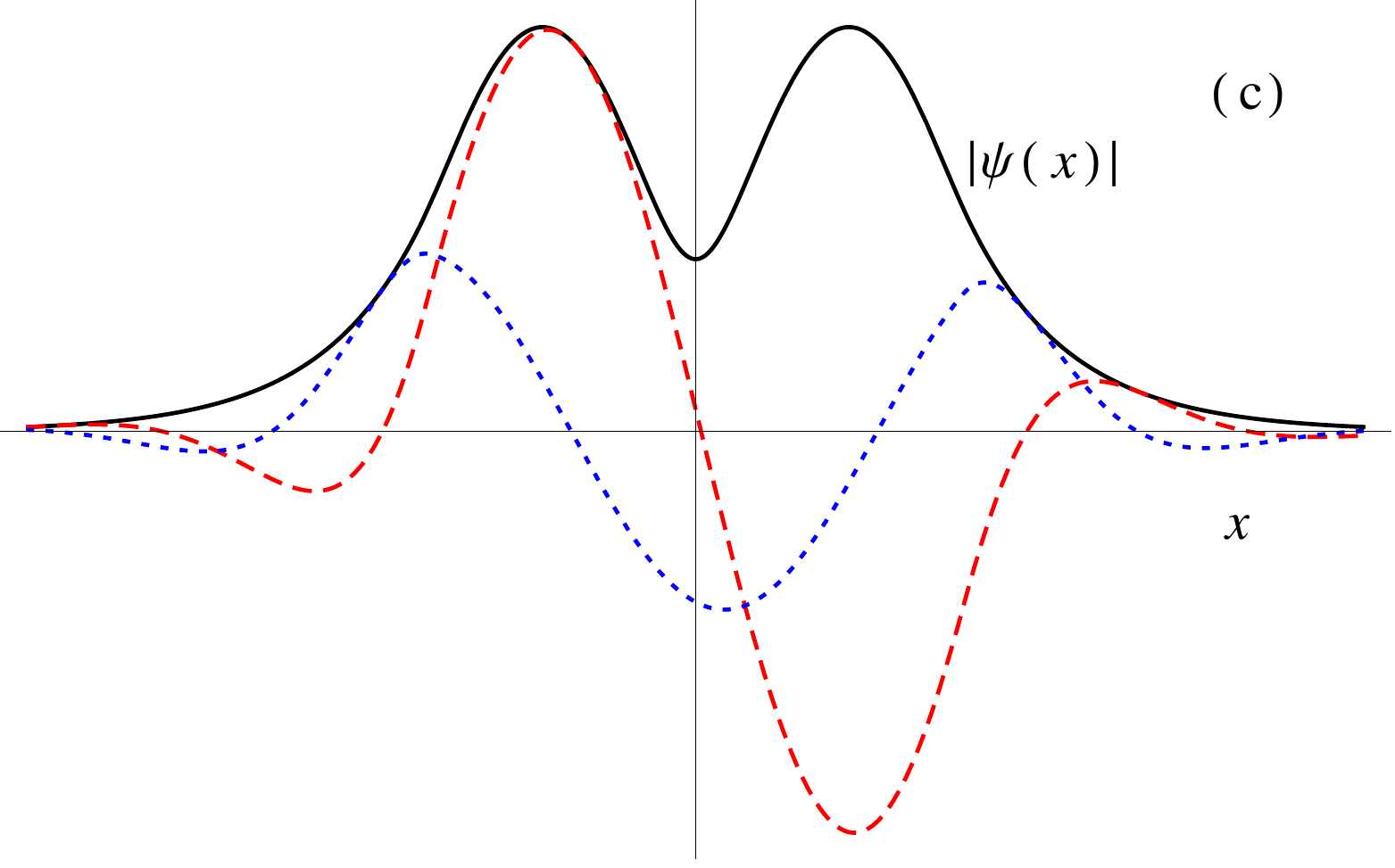}	
	\caption{Square well (11) potential for the  case: $V_0=0$,  $V_1=5, $, and $ a=8$. (a): Potential has only real discrete spectrum ( $E_1=0.4619, E_2=1.8754, E_3=4.3348, E_4=7.9631,....$), which is unbounded from above. (b): $|\psi(x)|$ (solid), $\Re(\psi(x))$ (dashed) and $\Im(\psi(x))$ (dotted) are plotted for $E=E_{1}$,  (c): The same as in (b) for $E=E_2$.}
\end{figure}

\begin{figure}[h]
	\centering
	\includegraphics[width=6 cm,height=6.cm]{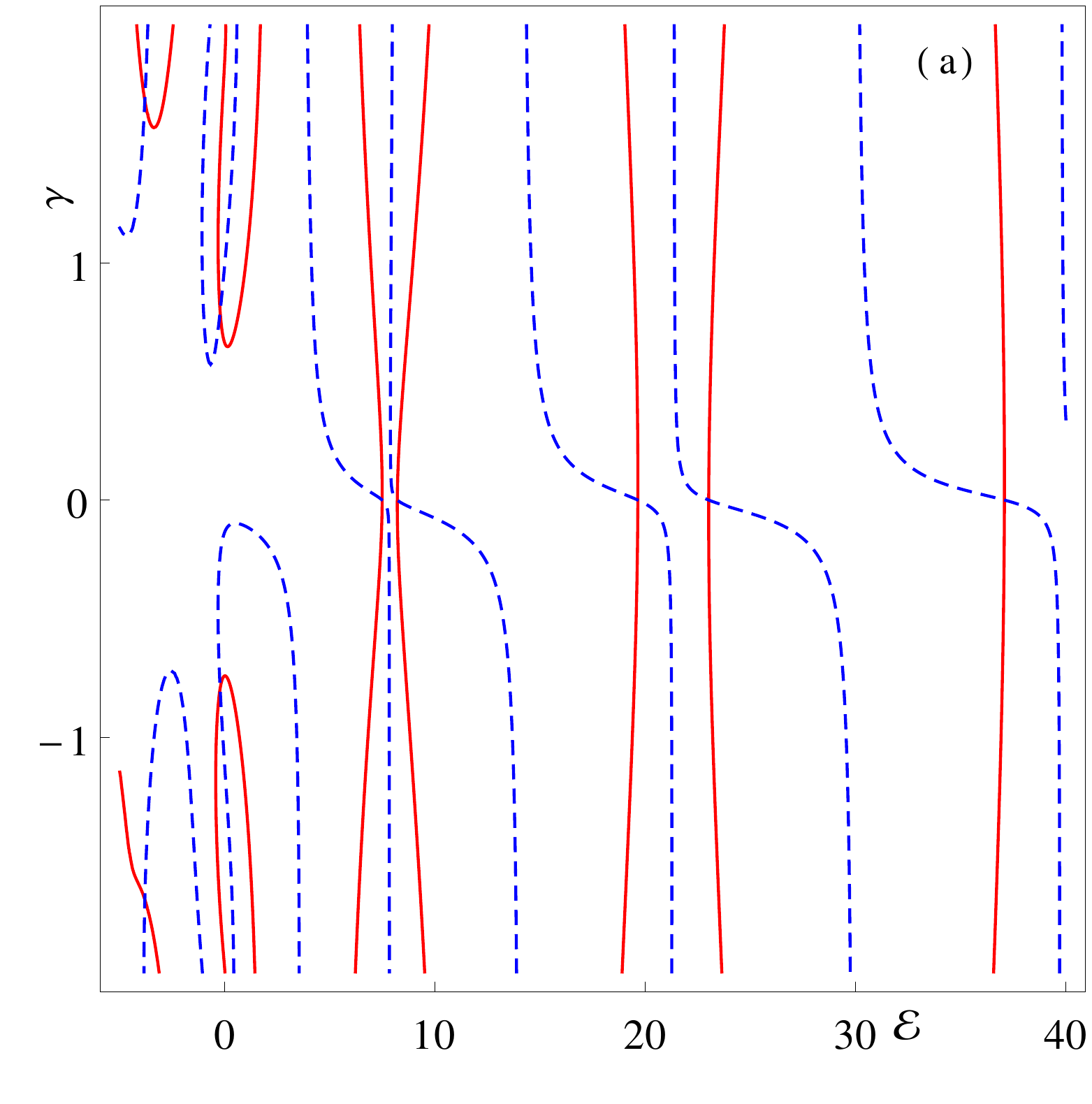}
	\includegraphics[width=5 cm,height=6.cm]{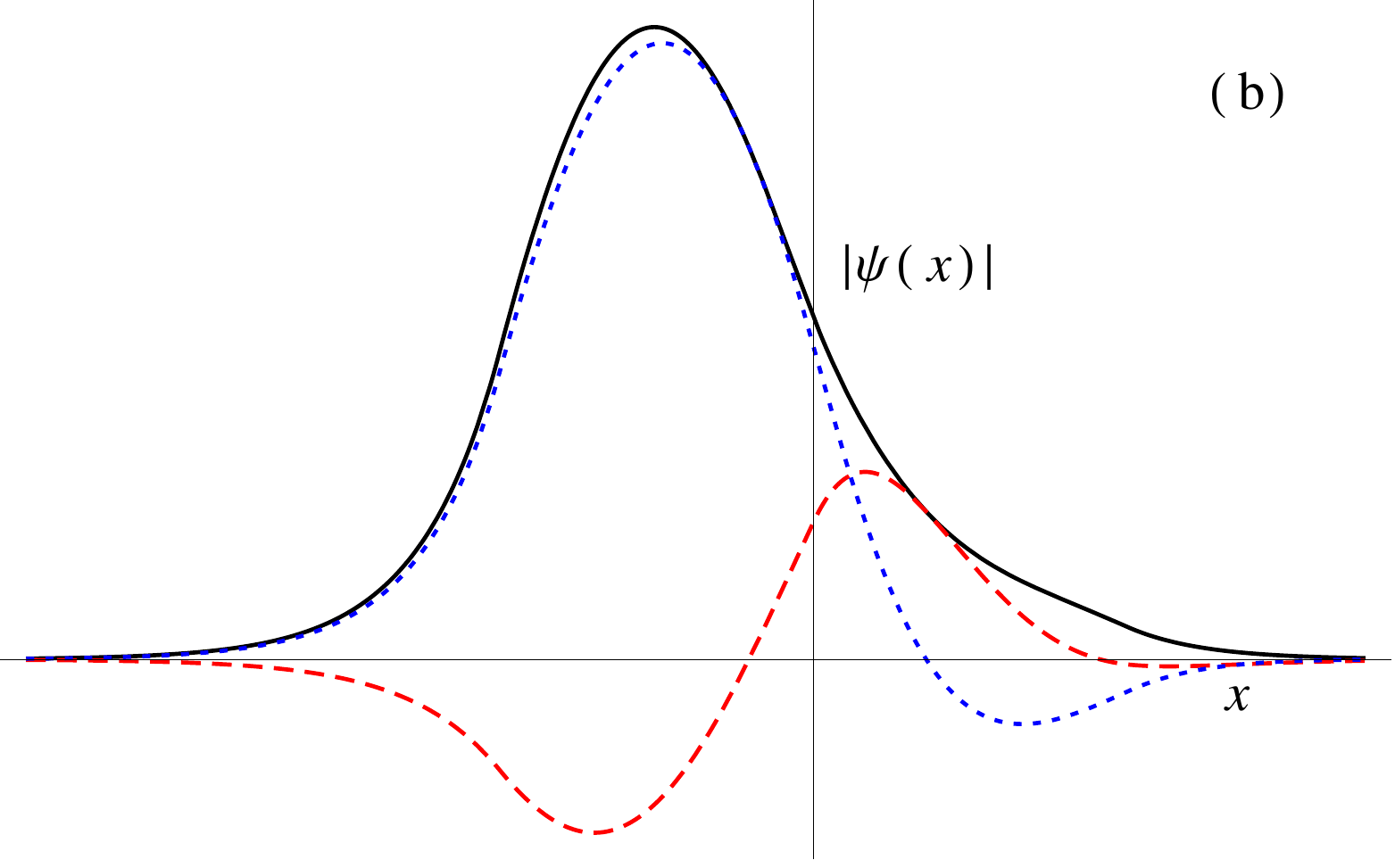}
	\includegraphics[width=5 cm,height=6.cm]{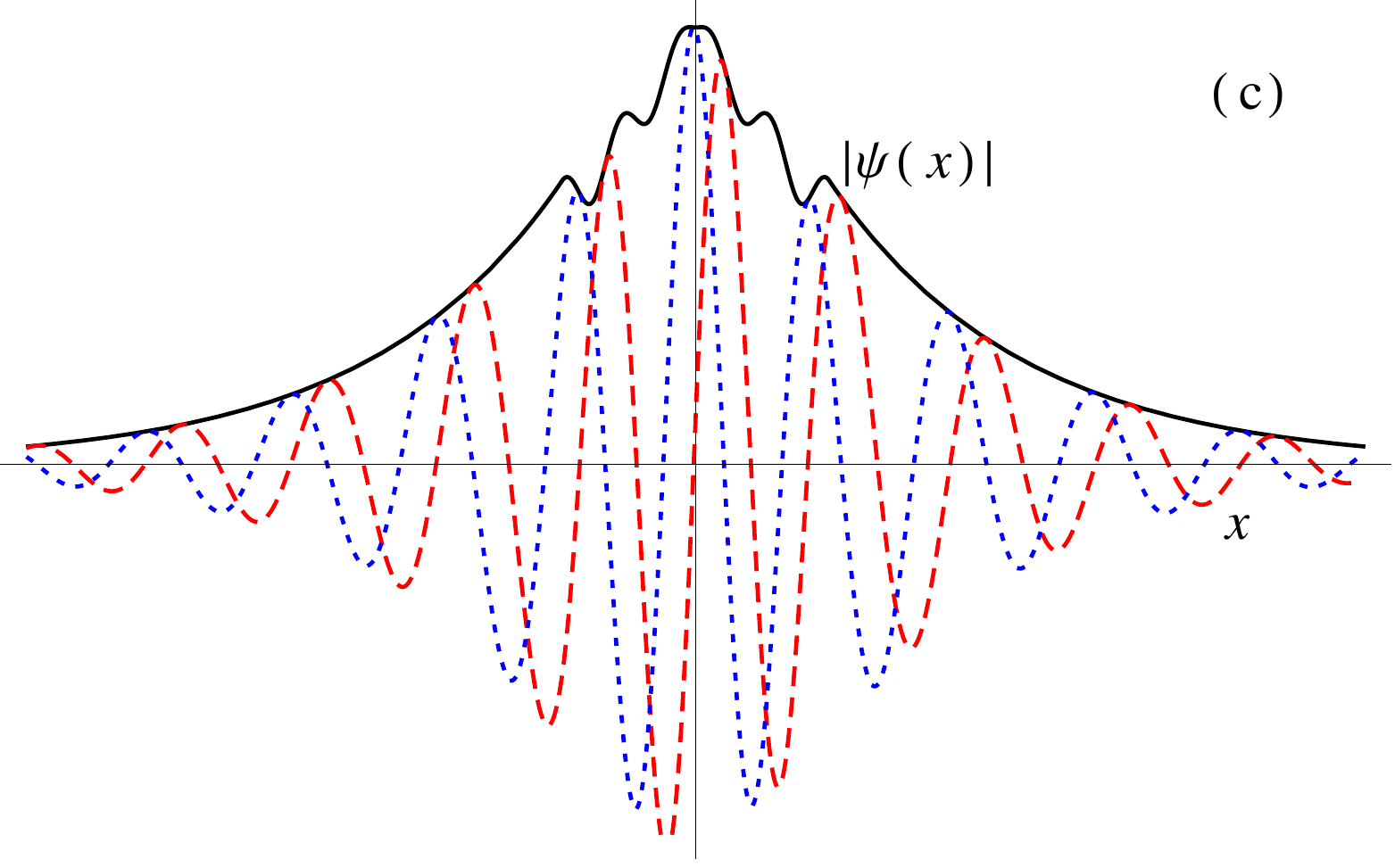}	
	\caption{Square well (11) potential for the  case: $V_1=2,V_0=5, a=2$. (a): The potential has both of complex conjugate pairs of eigenvalues ($E_{1\pm}=-3.8081\pm 1.6840i, E_2=-0.2267\pm 0.7944i$) and real discrete spectrum $(E_3=7.4807, E_4=8.2005, E_5=19.6416, E_5=23.0117, E_6=37.0765....$). If $V_1$ is increased, the lower real eigenvalues
		start disappearing and they convert to CCPEs. When $V_1\ge V_0$,  the real discrete spectrum is null.  (b):  $|\psi(x)|$ (solid), $\Re(\psi(x))$ (dashed) and $\Im(\psi(x))$ (dotted) are plotted for $E=E_1$. (c): The same as in (b) for $E=E_4$. Notice the for real energy $|\psi|$ peaks at $x=0$, and for complex energy it shifts from $x=0$. Eigenstates of CCPE satisfy the property (7). }
\end{figure}

\begin{figure}[h]
	\centering
	\includegraphics[width=6 cm,height=6.cm]{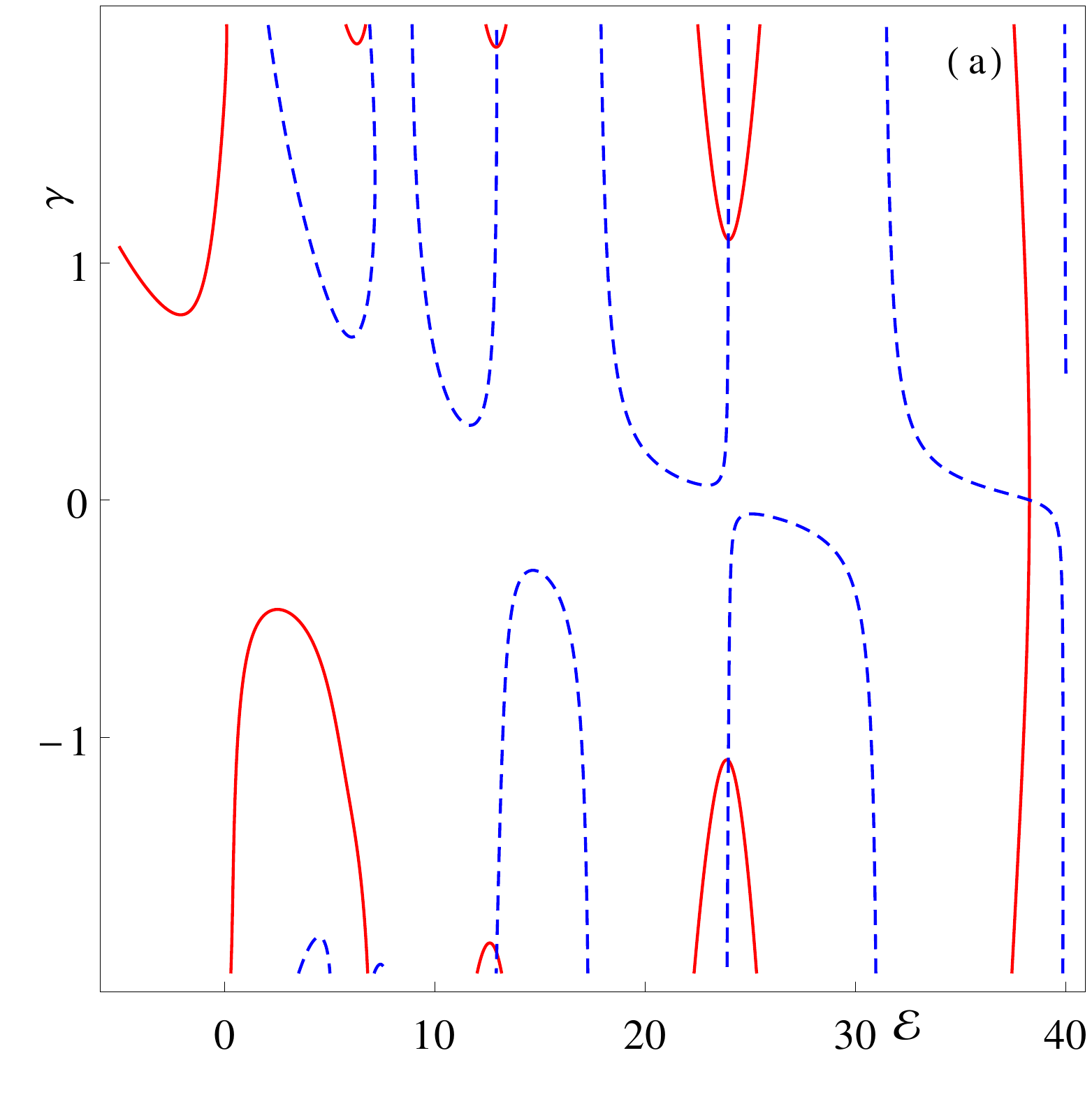}
	\includegraphics[width=5 cm,height=6.cm]{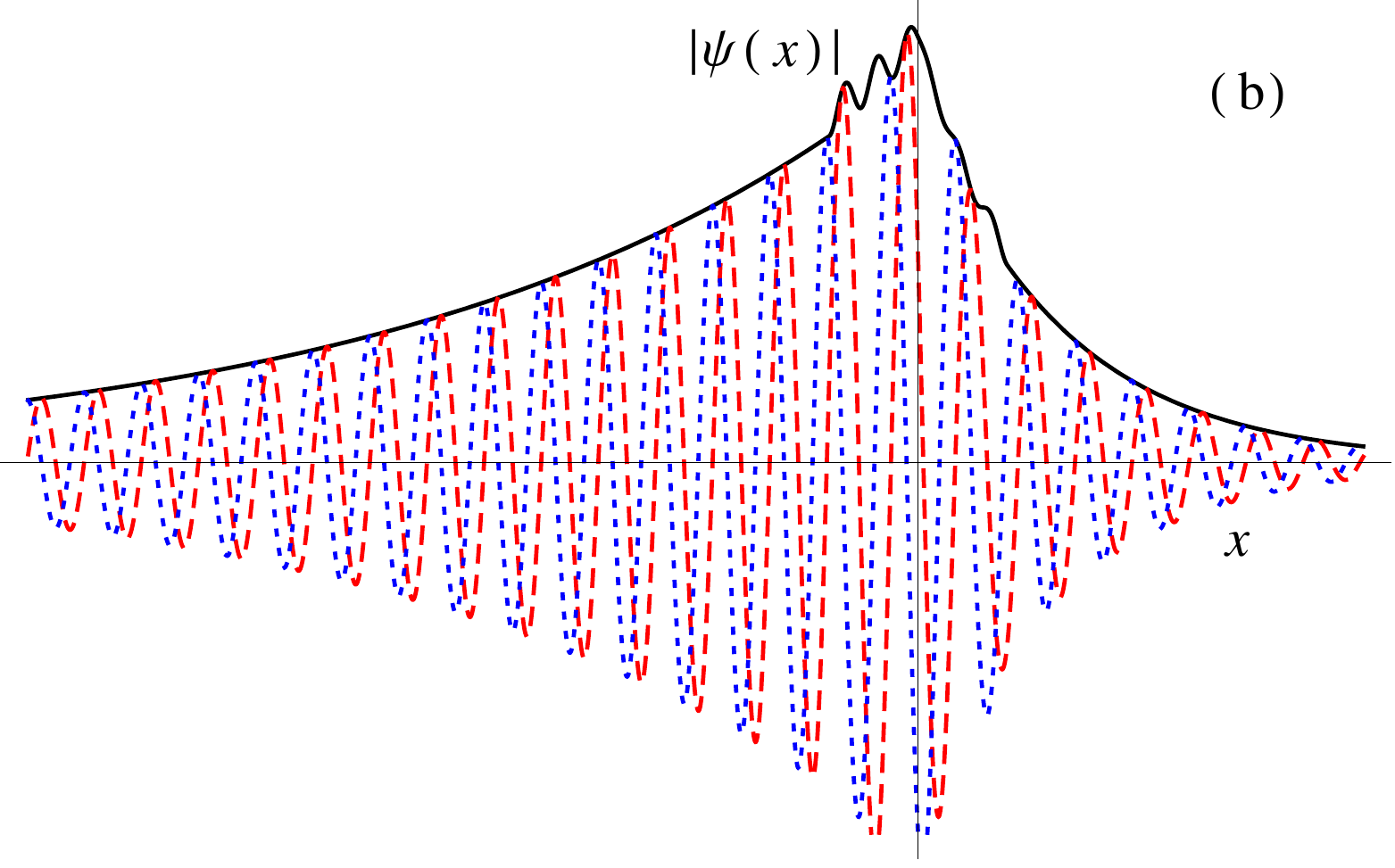}
	\includegraphics[width=5 cm,height=6.cm]{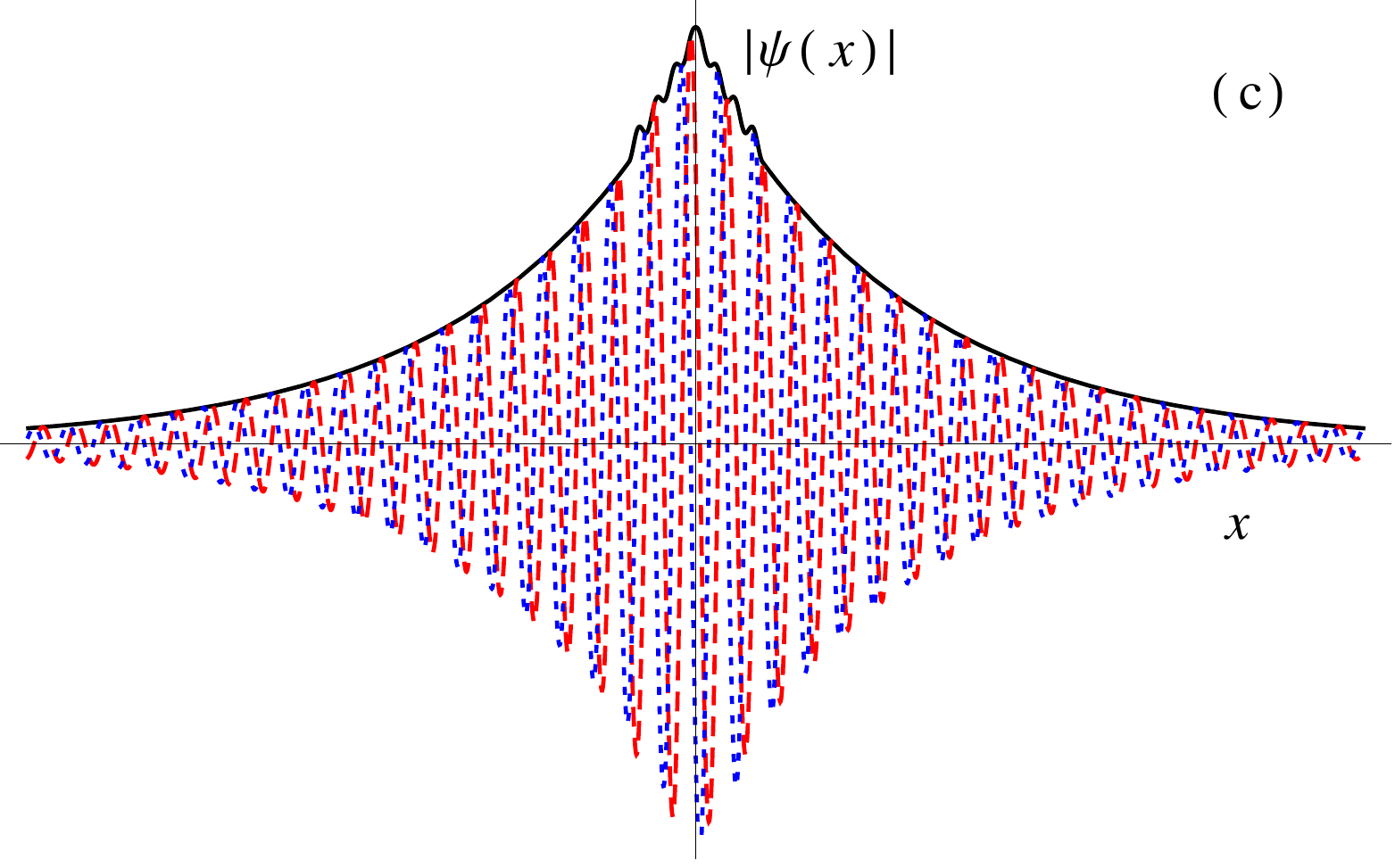}	
	\caption{Square well (11) potential for the  case: $V_1=2,V_0=-5, a=2$. (a) The potential has both of  complex conjugate pairs of eigenvalues ($ E_{1\pm}=12.9285 \pm i 1.9100, E_{2\pm}=23.9465\pm 1.0938i,$) and real discrete spectrum ($E_3=38.2666,...$) depicted by intersection of contours. (b): $|\psi(x)|$ (solid), $\Re(\psi(x))$ (dashed) and $\Im(\psi(x))$ (dotted) are plotted for $E=E_1$. (c): The same as in (b) for $E=E_2$. Notice the for real energy $|\psi|$ peaks at $x=0$, and, for complex energy it shifts from $x=0$. }
\end{figure}

Now we discuss the solved example of complex PT-symmetric Rosen-Morse potential [6,8]
\begin{equation}
V(x)=-s(s+1) \mbox{sech}^2x+2ic \tanh x;  ~~s \in \Re^+
\end{equation}
Its real, discrete energy spectrum is, 
\begin{equation} 
E_n=-(n-s)^2+\frac{c^2}{(n-s)^2},\quad n=0,1,2,3,...<s,
\end{equation}
with energy eigenstates as,
\begin{equation}
\psi(x)=A ~ \mbox{sech}^{(s-n)}x ~ e^{-icx/(s-n)} P^{s-n+ic/(s-n),s-n-ic/(s-n)}_n(\tanh x),
\end{equation}
where $P^{\alpha,\beta}_n(z)$ are the Jacobi's $n$ degree polynomials of $z$. One interesting feature of (17) is that by making $c$ larger the negative energies can be pushed to positive energies. A large value of $c$ makes the eigenfunction $\psi(x)$ more oscillatory. Even if $c$ is small, the closeness of $s$ to an integer brings in the same effect (See fig 6(d)). For instance for $s=3.2$, there will be four discrete energy states with eigenvalues as $E_0=-10.14,, E_1=-4.63,, E_2=-0.74, E_3=24.96.$, and in fig 6, we have plotted the energy eigen states ($|\psi|$ and $\Re(\psi(x))$ ) correspoding to these four discrete energies. Earlier these have been welcomed as normal real discrete energies of a complex PT-symmetric potential.  However, such a large and positive value for the last eigenvalue would attract ones attention, specially after the discovery of relfectionless states [14], one dimensional version of von Neumann's bound state embedded in positive energy continuum [13] and spectral singularity  (discrete positive energy scattering) state [16].
\begin{figure}[h]
	\centering
	\includegraphics[width=2 cm,height=3.cm]{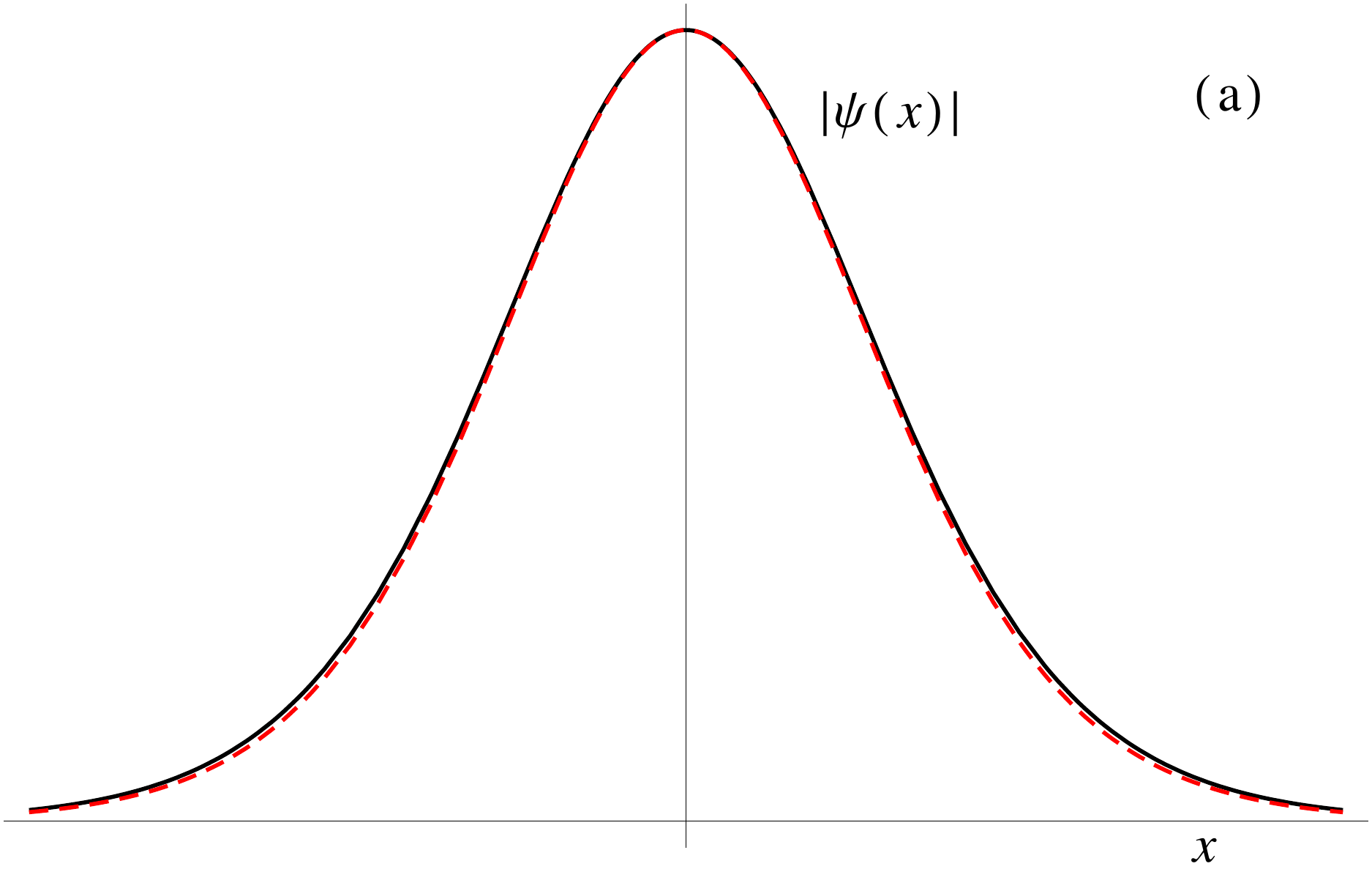}
	\includegraphics[width=3 cm,height=3.cm]{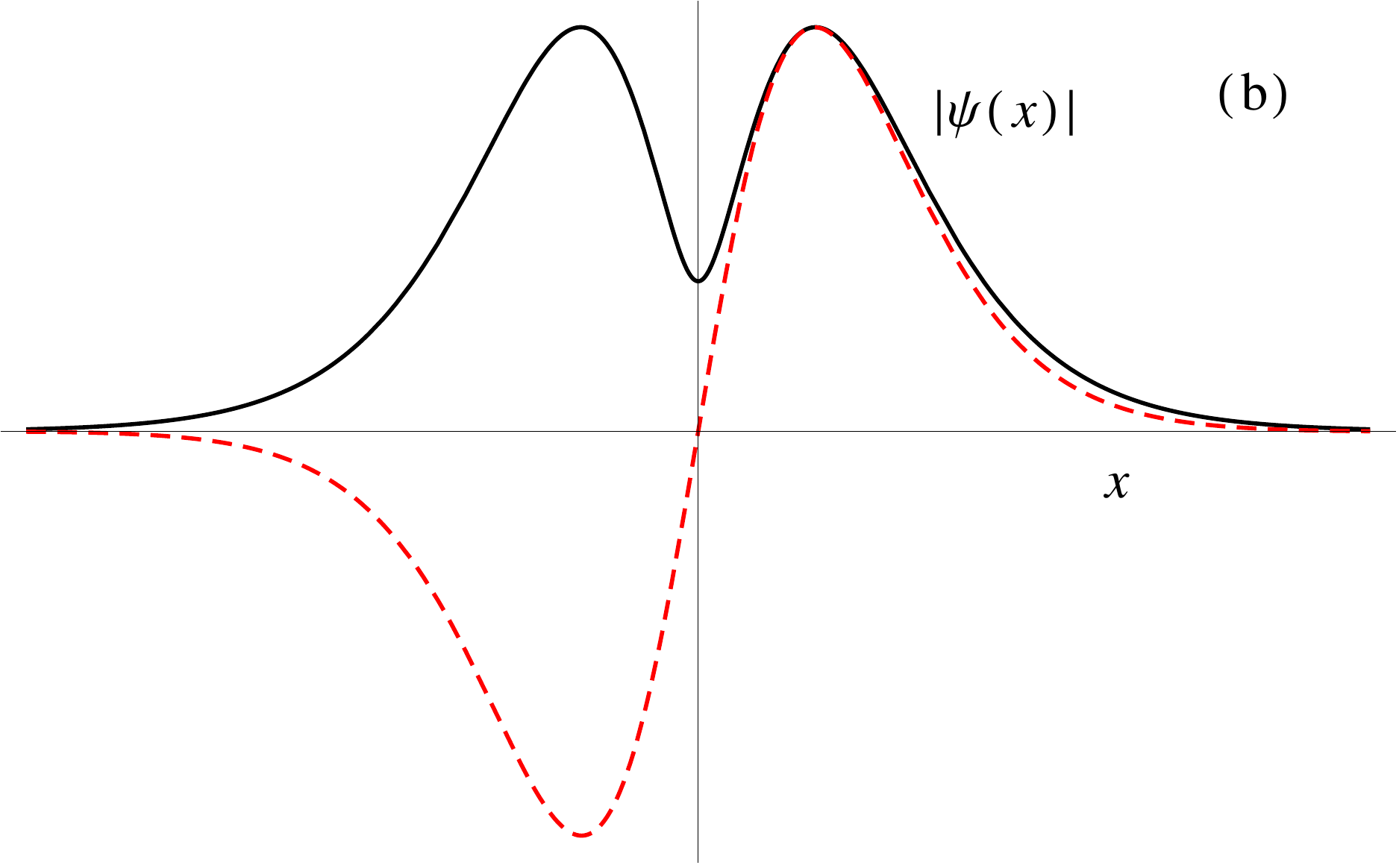}
	\includegraphics[width=4 cm,height=3.cm]{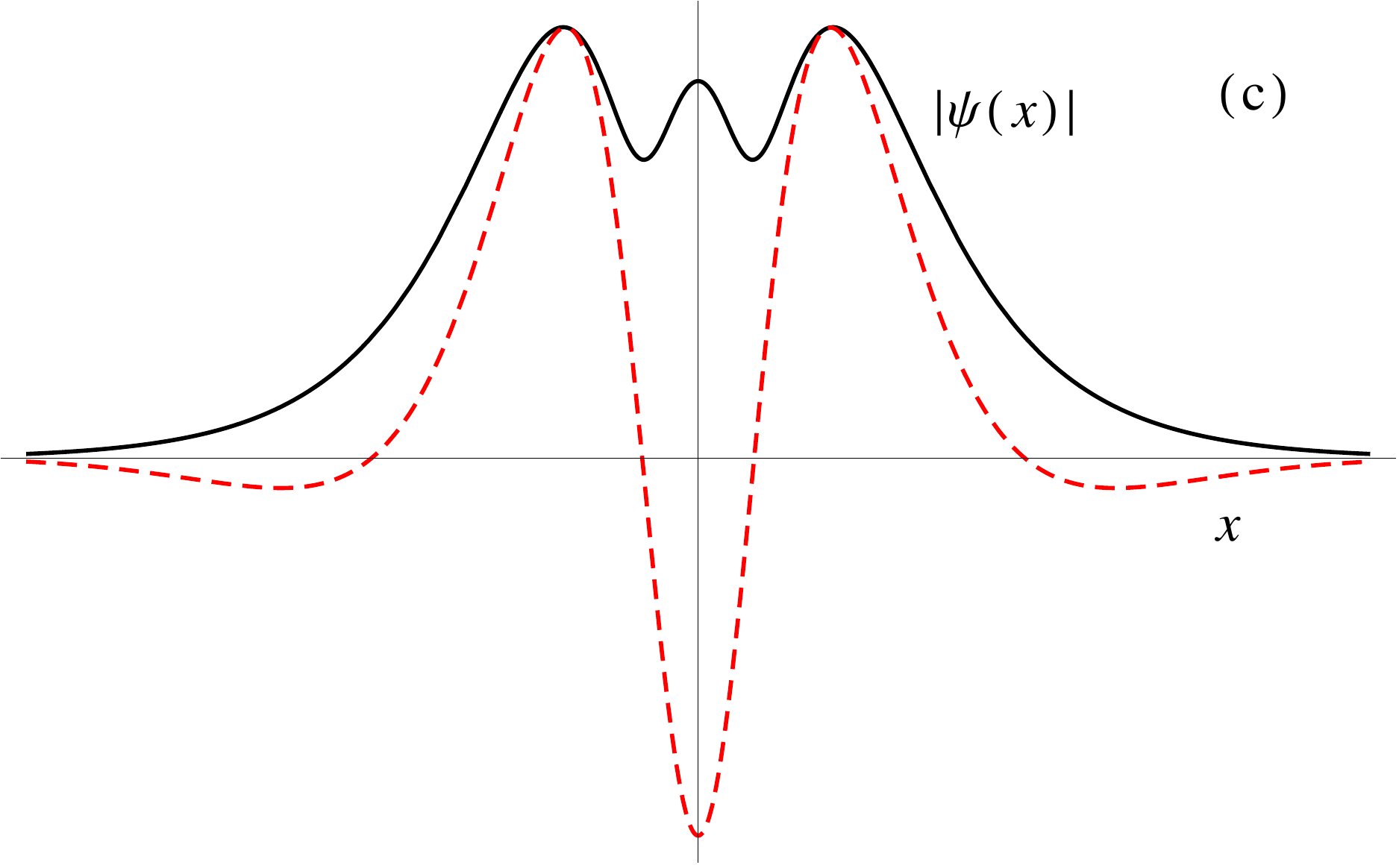}	
	\includegraphics[width=6 cm,height=3.cm]{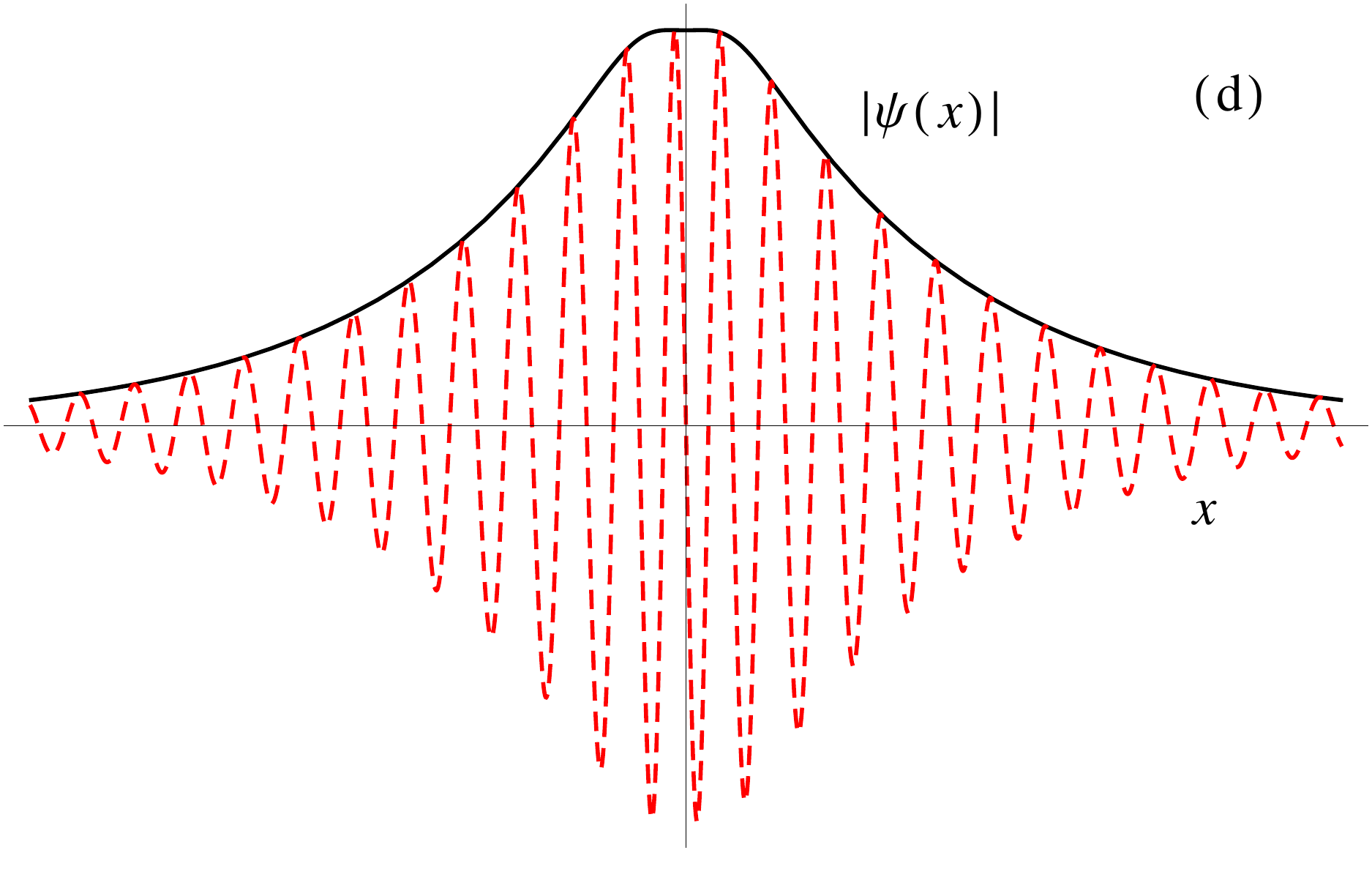}
	\caption{$|\psi|$ (solid) and $\Re(\psi(x))$ (dashed) of first four eigenstates (corresponding energy eigen values are $E_0=-10.14,, E_1=-4.63, E_2=-0.74, $ and $ E_3=24.96.$) of the complex PT-symmetric  potential (16) for the case: $s=3.2$ and $c=1$.  Notice the distinctive feature of fast oscillations in  the last state with positive energy eigenvalue.}
\end{figure}

According to Milne [11,12], a solution of Schr{\"o}dinger equation (1) can be written as $\psi(x)={\cal A}(x)e^{i{\cal S}(x)}$, it is as though the particle is in a  potential,
\begin{equation}
V(x)-E=\frac{\hbar^2}{2\mu}\left [\frac{{\cal A}''(x)}{{\cal A}(x)}-\frac{C^2}{{\cal A}^4(x)}\right], \quad {\cal S}'= \frac{C}{{\cal A}^2},
\end{equation}
where $C$ is arbitrary constant and energy $E$ is to be fixed. Interestingly, these wave solutions carry the current density $J$ as
\begin{eqnarray}
J&=&\frac{\hbar}{2i\mu}[ \psi^*(x) \psi'(x)-\psi(x) \psi^{*'}(x)]=\frac{\hbar}{\mu}{\cal A}^2 {\cal S}'= \frac{\hbar}{\mu} [\psi_r(x) \psi'_i(x)-\psi_i (x) \psi'_r(x)],
\end{eqnarray}
From Eq. (19), it correctly follows that $J$ is independent of $x$.
On the one hand, by choosing symmetric and asymptotically vanishing functions ${\cal A}(x)$ one can 
get the scattering states which carry unit current density, they travel in one direction without getting reflected. On the other hand. the real and imaginary parts of these eigenfunctions have been revealed [13] to give rise to degeneracy in one dimension at an energy $E$. See for example, eigenstates :$\frac{e^{i(x+x^3/3)}}{\sqrt{1+x^2}},\frac{e^{i \sinh x}}{\sqrt{\cosh x}}$ [13] and  $\frac{e^{\frac{i}{2} \sinh x}}{\sqrt{\cosh x}}$ [14]. These are also interesting instances of reflectionless states above their respective bottomless barrier or double barriers. These reflectionless states are akin to one dimensional version of  von Neumann states embedded in positive energy continuum [13]. These reflectionless states though vanish asymptotically yet they carry unit flux.

On the contrary, the positive  discrete energy states discussed here are like $\psi(x \sim - \infty)=A e^{(\alpha + i\beta)x}$ and $\psi(x \sim \infty)= B e^{(-\alpha+i\beta)x}$. For example, the eigenstate $\psi(x)=Ae^{ix} \mbox{sech}x$ (18)  of the PT-symmetric Rosen-Morse potential $V(x)=-2 \mbox{sech}^2x-2i\tanh x$ (16), has the position dependent current density  $J=A' \mbox{sech}^2x$ which also vanishes asymptotically. At an energy slighty different from their discrete energy eigenvalue, these states may diverge at large distances or if they converge, they become non-differentiable at the joints of the potential. 

We find that PT-symmetric potentials with imaginary asymptotic saturation are strictly devoid of scattering states, they not only vanish asymptotically but also their current density is position dependent which vanishes asymptotically. Earlier studies [8,9] have ignored this point and even found reflection and transmission amplitudes which being inconsistent do not degenerate to those [15] of Hermitian Rosen Morse potential (15) when $c$ is set equal to zero or made imaginary. Some of these potentials (4,8,11)  can have real discrete spectrum with or without complex conjugate pairs of eigenvalues, earlier works have overlooked the existence of CCPEs in them. 

For complex PT-symmetric scattering potentials which vanish asymptotically [4-7], a very interesting type of real discrete positive energy $E_*$ called spectral singularity has been discovered [16]. At this special energy the transmission and reflection probabilities become infinite and the scattering state becomes plane wave as $e^{\pm ikx}$. Recently, it has been conjectured [17] that such  a fixed potential has at most one SS and if it exists, it sets the upper (or rough upper) bound to the real part of complex conjugate pairs of eigenvalues. Real discrete eigenstates of the present class of potentials oscillate but they vanish asymptotically so they cannot represent the spectral singularity state [16]. Recently a non-Hermitian and non-PT-symmetric version of complex Morse potential [18] has been employed to investigate the phenomenon of SS and coherent perfect absorption.

We find that $|\psi(x)|$ for real discrete eigenstates for this class of PT-symmetric potentials (4,8,11,16) are again integrable, symmetric and node-less  as  conjectured in Ref. [19] for other kinds of PT symmetric potentials [1-7]. For   CCPEs ($E_{\pm}$) the eigenstates $\psi_{\pm}$ satisfy  the property (7) (see Figs. 1(b,c)).
It is worthwhile to check that this property holds in general for complex PT-symmetric potentials having CCPEs, where $|\psi_{\pm}(x)|$ is integrable and asymmetric function peaking on the left/right of $x=0$ (see Figs. 2(b), 4(b), 5(b)). This interesting discriminatory feature of spontaneously broken PT-symmetry has been overlooked so far.

We have brought attention to the fast oscillatory behaviour of higher energy real discrete states which is likely to be confused with similar behaviour of  quantal states which are reflectionless, one dimensional version of von Neumann's boundstate embedded in positive energy continuum and the special positive enegy discrete state called spectral singularity. It will not be surprising that the fast oscillatory behaviour of these higher discrete states and the interesting invariance property  of $|\psi_{\pm}(\pm x)|$  proposed here in Eq. (7) for broken PT-symmetry, in general, could be harnessed experimentally in optics where complex PT-symmetric mediums have already thrown novel effects and surprises [20]. Moreover, new exactly solvable models presented here would attract attention in this regard.

\end{document}